\pdfoutput=1

\newcommand{\ybco}{YBa$_2$Cu$_3$O$_{6 + y}$}

\documentclass[twocolumn,showpacs,preprintnumbers,amsmath,amssymb,superscriptaddress,prl]{revtex4}

\usepackage{graphicx}% Include figure files
\usepackage{dcolumn}% Align table columns on decimal point
\usepackage{bm}% bold math

\begin{document}

%\preprint{To appear in Physical Review Letters, accepted 11 October 2007. (arXiv:cond--mat/0509223)}

\title{Superfluid Density in a Highly Underdoped \ybco\ Superconductor}% Force line breaks with \\
\author{D.~M.~Broun}
\author{W.~A.~Huttema}
\author{P.~J.~Turner}
\affiliation{Department of Physics, Simon Fraser University, Burnaby, BC, V5A 1S6, Canada}
\author{S.~\"Ozcan}
\author{B.~Morgan}
\affiliation{Cavendish Laboratory, Madingley Road, Cambridge, CB3 0HE, United Kingdom}
\author{Ruixing Liang}
\author{W.~N.~Hardy}
\author{D.~A.~Bonn}
\affiliation{Department of Physics and Astronomy, University of British Columbia, Vancouver, BC, V6T 1Z1, Canada}%

\date{\today}% It is always \today, today,

\begin{abstract}
The superfluid density $\rho_{\rm s}(T) \equiv 1/\lambda^2(T)$ has been measured at 2.64~GHz in highly underdoped \ybco, at 37 dopings with $T_{\rm c}$ between 3~K and 17~K.  Within limits set by the transition width $\Delta T_{\rm c} \approx 0.4$~K, $\rho _{\rm s}(T)$ shows no evidence of 
critical fluctuations as $T \to T_{\rm c}$, with a mean-field-like transition and no indication of vortex unbinding.  Instead, we propose that $\rho_{\rm s}$ displays the behaviour expected for a quantum phase transition in the $(3 + 1)$-dimensional $XY$ universality class, with $\rho_{\rm s0} \propto (p - p_{\rm c})$, $T_{\rm c} \propto (p - p_{\rm c})^{1/2}$ and $\rho _{\rm s}(T) \propto (T_{\rm c} - T)^{1}$ as $T \to T_{\rm c}$.
\end{abstract}

\maketitle

Current research on high temperature superconductivity focuses on the underdoped cuprates, in
a region of the phase diagram where $d$-wave superconductivity gives way to
antiferromagnetism \cite{orenstein00}.  One proposal for this regime is that at
temperatures up to about 100~K above $T_{\rm c}$, superconductivity persists locally, with
long-range phase coherence suppressed by fluctuations in the phase of the superconducting
order parameter \cite{emery95,franz01,herbut02, herbut02a,herbut04,herbut05,franz06}. Early results showing a
linear relation between $T_{\rm c}$ and the superfluid density $\rho_{\rm s}(T = 0)$ \cite{uemura89} provided
the original motivation for this point of view, suggesting that $T_{\rm c}$ is low in underdoped
materials because the phase stiffness is low.  Further support for this idea has come from
measurements showing a finite phase stiffness above $T_{\rm c}$ at terahertz
frequencies \cite{corson99}, and from experiments that appear to detect the phase-slip voltage
of thermally diffusing vortices in the normal state \cite{wang03}.  If the physics of the underdoped cuprates is indeed that of a fluctuating $d$-wave superconductor there should be a regime where quantum fluctuations come into play as $T_{\rm c}$ falls to zero with decreasing doping.  Here we test this idea with a detailed study of the doping dependence of the superfluid density in the vicinity of the critical doping for superconductivity.

 High homogeneity of $T_{\rm c}$ is particularly difficult to achieve in the underdoped cuprates, where the control parameter is chemical doping and the materials are well away from plateaus or turning points in $T_{\rm c}(y)$. The \ybco\ system has two advantages in this doping range: with careful work, there can be sufficient control of doping homogeneity to produce samples with sharp superconducting transitions \cite{liang02,liang06}; and the
process of CuO-chain ordering can be harnessed to provide continuous tunability of the carrier density in a
{\em single} sample, with {\em no} change in cation disorder \cite{hosseini04}. This is possible because the loosely held chain oxygen atoms  in these high quality samples remain mobile at
room temperature and gradual ordering into CuO-chain structures slowly pulls electrons from the CuO$_2$ planes, smoothly increasing hole doping over time \cite{zaanen88,veal90}. For this experiment, single crystals of \ybco\ were grown in barium zirconate crucibles and have high purity and low defect levels, with cation disorder at the $10^{-4}$ level \cite{liang98}.  A crystal 0.3~mm thick was cut and polished with Al$_2$O$_3$ abrasive into an ellipsoid of revolution about the sample $c$-axis, 0.35~mm in diameter. The oxygen content of the ellipsoid was adjusted to O$_{6.333}$ by annealing at 914$^\circ$C in flowing oxygen, followed by a homogenization anneal
in a sealed quartz ampoule at 570$^\circ$C and a quench to 0$^\circ$C.  At this point the sample was nonsuperconducting.  After allowing chain oxygen
order to develop at room temperature for three weeks, $T_{\rm c}$ was 3~K.  The sample was then further annealed at room temperature for six weeks under
hydrostatic pressure of $\sim30$~kbar, raising $T_{\rm c}$ to 17~K.  The sample was cooled to $-5^\circ$C, removed from the pressure cell, and then
stored at $-10^\circ$C to prevent the oxygen order from relaxing. All further manipulation of the ellipsoid was carried out in a refrigerated
glove box at temperatures less than \mbox{$-5^\circ$C}.  In between measurements of surface impedance, periods of controlled \emph{in-situ} annealing at  room temperature and ambient pressure were
used to generate a sequence of 36 dopings as $T_{\rm c}$ relaxed back to 3~K. We emphasize that no oxygen entered or left the sample during these annealing steps.  Subsequent reannealing under hydrostatic pressure, for a further six weeks, returned $T_{\rm c}$ to the starting value of 17~K, where the sample was remeasured to demonstrate the reversibility of the technique. 

\begin{figure}[t]
\begin{center}
\includegraphics[width=83mm]{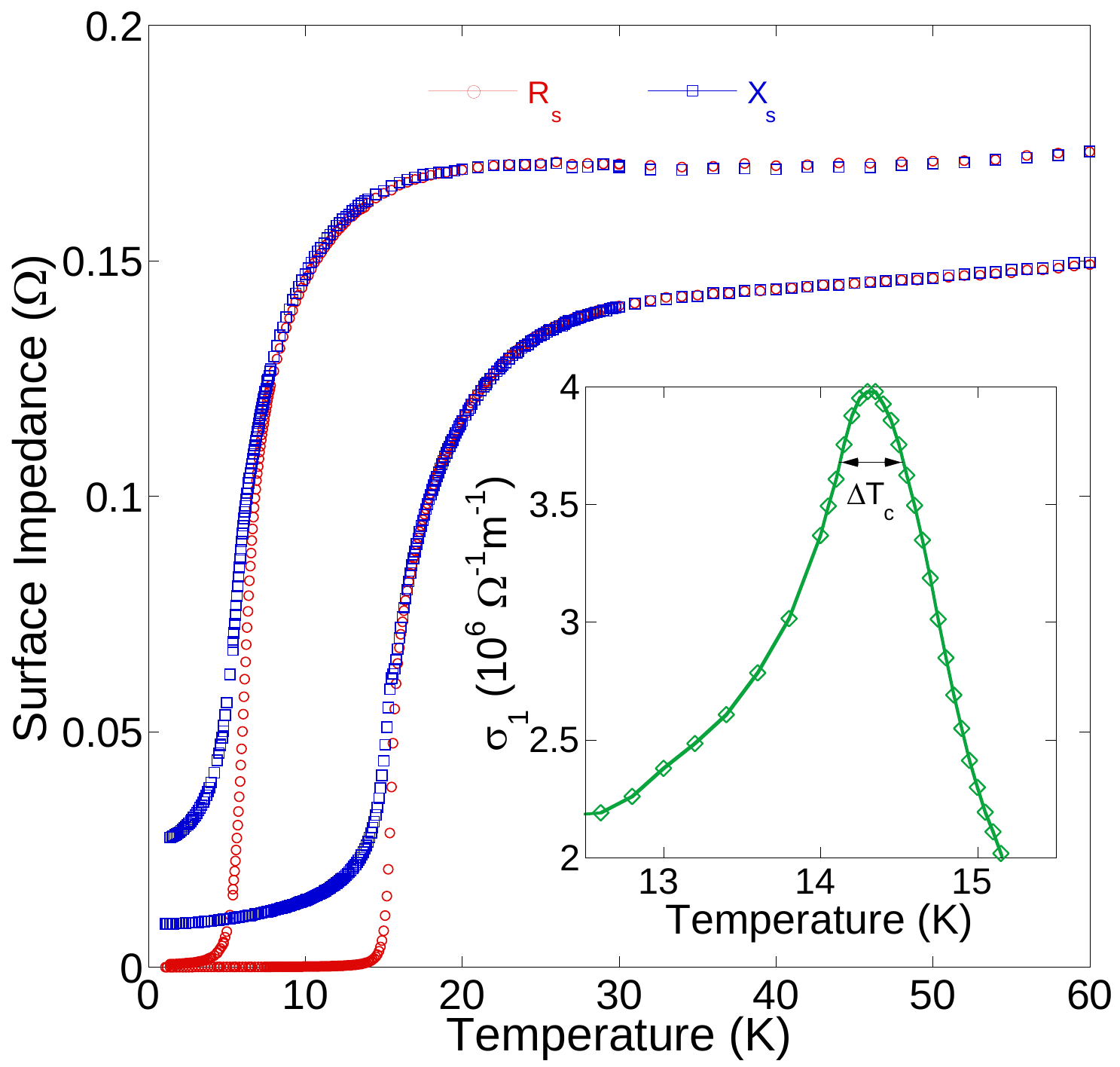}
\caption{(color online).  $ab$-plane surface impedance at 2.64~GHz for the \ybco\ ellipsoid, at two dopings.  $R_{\rm s}(T)$ is measured directly in the
experiment.  Absolute reactance is obtained by offsetting $\Delta X_{\rm s}(T)$ so that $R_{\rm s}$ and
$X_{\rm s}$ match in the normal state. Inset: a fluctuation peak in $\sigma_1(T)$ at one doping.  $\Delta T_{\rm c}$
is set to the difference between inflection points in $\sigma_1(T)$ on opposite sides of the
transition. } \label{fig1}
\end{center}
\end{figure}

\begin{figure}[t]
\includegraphics[width=83mm]{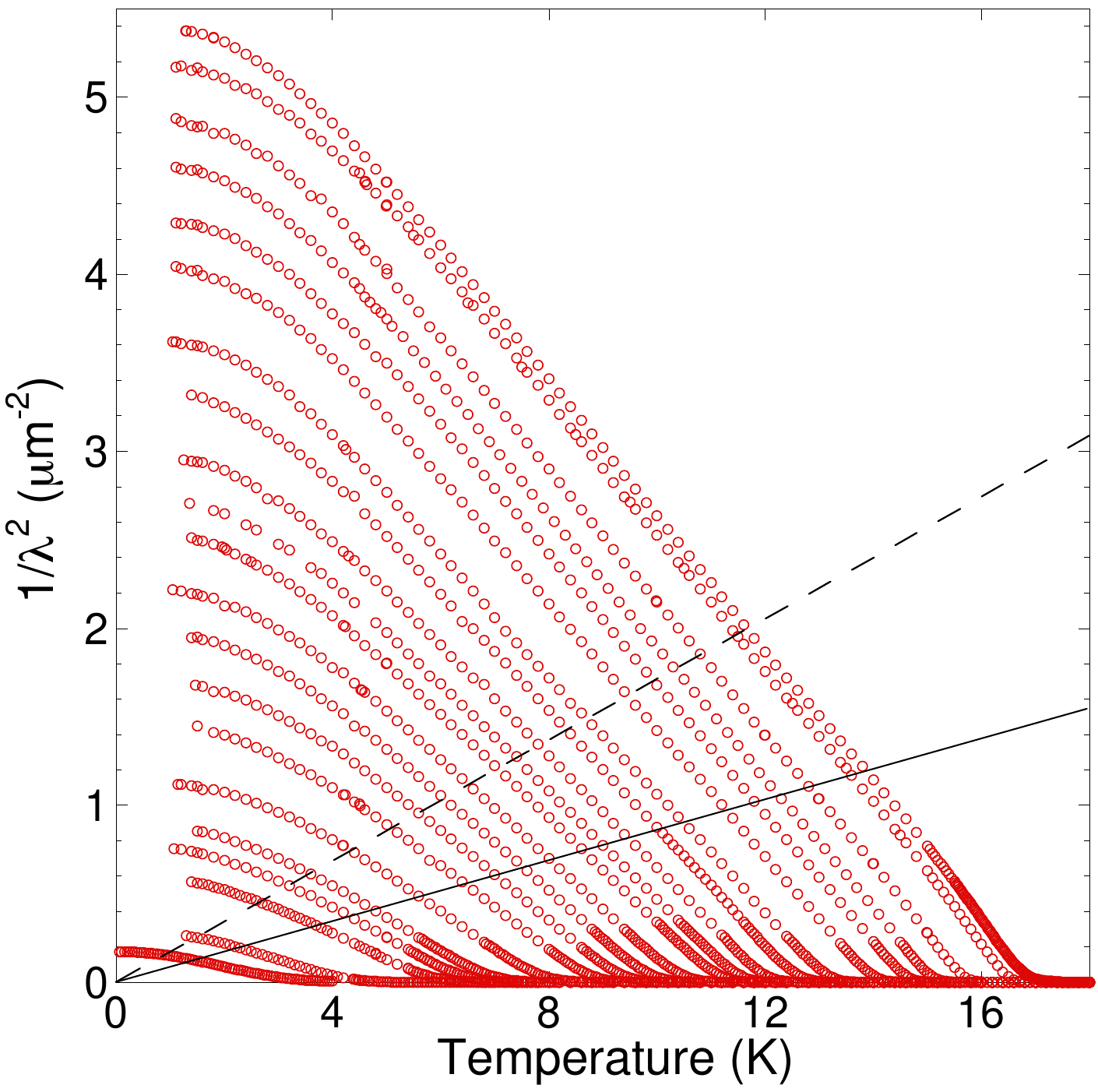}
\caption{\label{fig2} (color online).  $ab$-plane superfluid density $\rho_{\rm s}(T) = 1/\lambda^2(T)$  shown at 20 of the 37 dopings measured in this study.   Measurements were made starting in the most ordered state ($T_{\rm c} \approx
17$~K) followed by controlled oxygen annealing in small steps down to $T_{\rm c} \approx 3$~K. At the end of the experiment the sample was reordered and measured again to verify reproducibility.  Lines
mark where the vortex-unbinding transition should occur for a 2D superconductor.  The dashed
line corresponds to \mbox{$\rho_{\rm s}^{\rm 2D} \equiv  \hbar^2 d/4 k_{\rm B} e^2 \mu_0\lambda^2 = (2/\pi)T$} in each CuO$_2$ plane.  The
solid line shows $\rho_{\rm s}^{\rm 2D} = (2/\pi)T$ in each CuO$_2$ bilayer.  $\rho_{\rm s}(T)$ instead passes smoothly through this region.  While mean-field-like over most of the doping range, $\rho_{\rm s}(T)$ develops downwards curvature near $T_{\rm c}$ at the highest dopings, a possible indication of the onset of the 3D-$XY$ critical fluctuations.
}
\end{figure}

Measurements of the $ab$-plane surface impedance $Z_{\rm s} = R_{\rm s} + \textrm{i} X_{\rm s}$ were carried
out at 2.64~GHz by cavity perturbation, using a sapphire hot-finger to position the sample at
the $H$-field antinode of the TE$_{01\delta}$ mode of a rutile dielectric
resonator \cite{huttema05}. All data sets reported here were taken with the $c$-axis of the ellipsoid oriented along the microwave magnetic field $H_{\rm rf}$ to induce $ab$-plane screening currents. The surface impedance of the sample has been obtained from the measured cavity response using the cavity perturbation formula
$\Delta f_B(T) - 2\textrm{i}\Delta f_0(T) = \Gamma(R_{\rm s} + \textrm{i} \Delta X_{\rm s})$, where $\Delta f_B(T)$ is the change in bandwidth of the TE$_{01\delta}$
mode upon inserting the sample into the cavity,  and $\Delta f_0(T)$ is the shift in resonant frequency upon warming the sample from base temperature to $T$.
$\Gamma$ is a scale factor that applies to the data set as a whole, and is empirically determined using a Pb--Sn replica sample to an accuracy of 2.5\%.  
The absolute surface reactance is set in the usual way by matching $R_{\rm s}$ and $X_{\rm s}$ in the normal state, where we expect the imaginary part of the microwave conductivity to
be very small.   This is illustrated in Fig.~\ref{fig1}, which shows the surface impedance $Z_{\rm s} = R_{\rm s} + \textrm{i} X_{\rm s}$ of the
ellipsoid at two of the dopings.  The rounded shoulders in $Z_{\rm s}(T)$ are the result of fluctuations.  This is seen more clearly in the microwave conductivity  $\sigma = \sigma_1 - \textrm{i} \sigma_2$, which is obtained using the local-limit expression $\sigma = \textrm{i} \omega \mu_0/ Z_{\rm s}^2$. The inset of Fig.~\ref{fig1} shows $\sigma_1(T)$ at one of the dopings, revealing a narrow fluctuation peak at $T_{\rm c}$. In a homogeneous system, $\sigma_1(T)$ is expected to have a sharp cusp at $T_{\rm c}$ \cite{fisher91}. Rounding of the fluctuation peak, arising from the macroscopic dopant inhomogeneity present in all real samples, is used to define $\Delta T_{\rm c}$.

\begin{figure}[t]
\includegraphics[width=80mm]{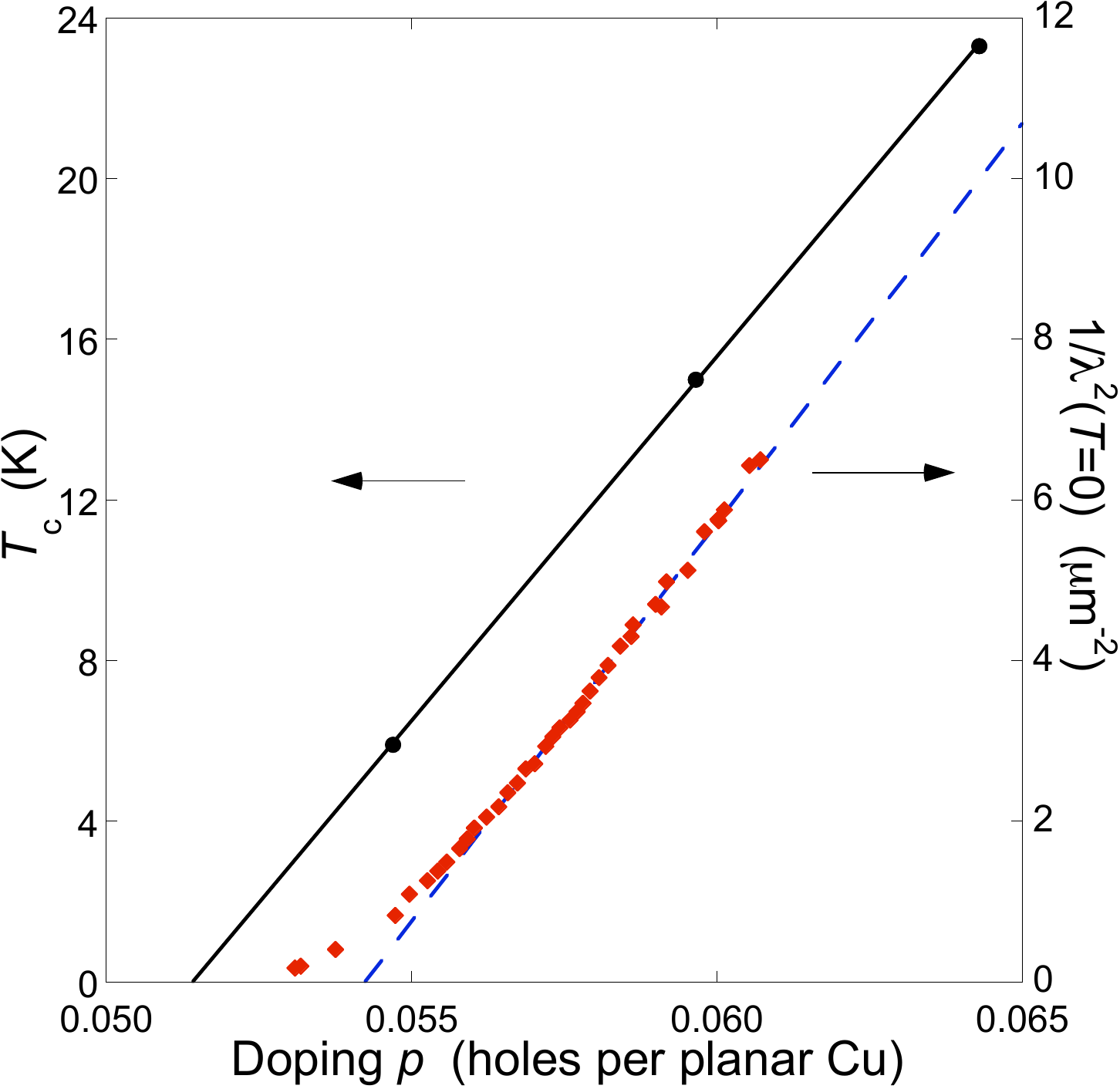}
\caption{\label{fig3} (color online).  Our $\rho_{s0}(p)$ data (solid diamonds, right-hand scale) and $T_{\rm c}(p)$ data from Ref.~\onlinecite{liang06} (solid circles, left-hand scale).  $\rho_{s0} \equiv
1/\lambda^2(T\rightarrow 0)$ is obtained by linear extrapolation to $T =
0$. Hole doping for the superfluid density data is determined from our values of $T_{\rm c}$, using a linear fit and extrapolation of the $T_{\rm c}(p)$ data from Ref.~\onlinecite{liang06} (solid line).   To the extent that this extrapolation is valid,  $\rho_{s0}(p) \sim (p - p_{\rm c})^2$ at the onset of superconductivity,  before crossing over to a linear doping dependence.  The dashed line is a linear fit to $\rho_{s0}(p)$ at higher doping.}
\end{figure}

The superfluid density is given by $\rho_{\rm s} \equiv 1/\lambda^2 = \omega \mu_0 \sigma_2$.  Figure~\ref{fig2} shows $\rho_{\rm s}(T)$ at 20 of the 37 dopings.  The most striking feature of the data is the wide range of linear temperature dependence, extending from close to  $T_{\rm c}$ down to $T\approx 4$~K.  Below 4~K $\rho_{\rm s}(T)$ crosses over to an accurately quadratic temperature dependence.  Such behaviour is well established in \ybco\ at higher dopings and is consistent with $d$-wave superconductivity in the presence of a small density of pair-breaking defects \cite{bonn94, hirschfeld93}.  One unusual feature of the data is the nature of the thermal transitions, which, within limits set by $\Delta T_{\rm c}$ appear mean-field-like.  At optimal doping \ybco\ is the most three dimensional cuprate, with $\lambda_c^2(T\rightarrow0)/\lambda_{ab}^2(T\rightarrow0)
\approx 50$ \cite{peregbarnea04}.  Its critical behaviour has been firmly established to be in the 3D-$XY$ universality class \cite{kamal94,junod00,meingast01}.  In the doping range explored in this paper, \ybco\ is highly anisotropic, with \mbox{$\lambda_c^2(T\rightarrow0)/\lambda_{ab}^2(T\rightarrow0)
\approx 10,000$} \cite{hosseini04}.  In these circumstances, one would anticipate fluctuations in adjacent layers to be uncorrelated, and a Kosterlitz--Thouless--Berezinsky (KTB) vortex unbinding transition \cite{ioffe02,herbut04} should occur when the 2D phase stiffness in a layer of
thickness $d$, \mbox{$\rho_{\rm s}^{\rm 2D}(T) \equiv \hbar^2 d/4 k_{\rm B} e^2 \mu_0\lambda^2(T)$}, falls to
$(2/\pi)T$.  This defines minimum superfluid densities for isolated planes and CuO$_2$ bilayers, shown in Fig.~\ref{fig2} by the dashed and solid lines respectively.
$\rho_{\rm s}(T)$ instead passes smoothly through these lines, with no indication of vortex unbinding.  Surprisingly, this implies that fluctuations remain correlated over many unit cells in the $c$ direction.  Recent work 
on \ybco\ thin films supports this, showing that the KTB transition does occur but that the effective thickness for fluctuations is the film thickness \cite{zuev04}.  At the highest dopings in our experiment, $\rho_{\rm s}(T)$ develops slight downward curvature near $T_{\rm c}$, possibly indicating the emergence of 3D-$XY$ criticality. 

\begin{figure}[t]
\includegraphics[width=80mm]{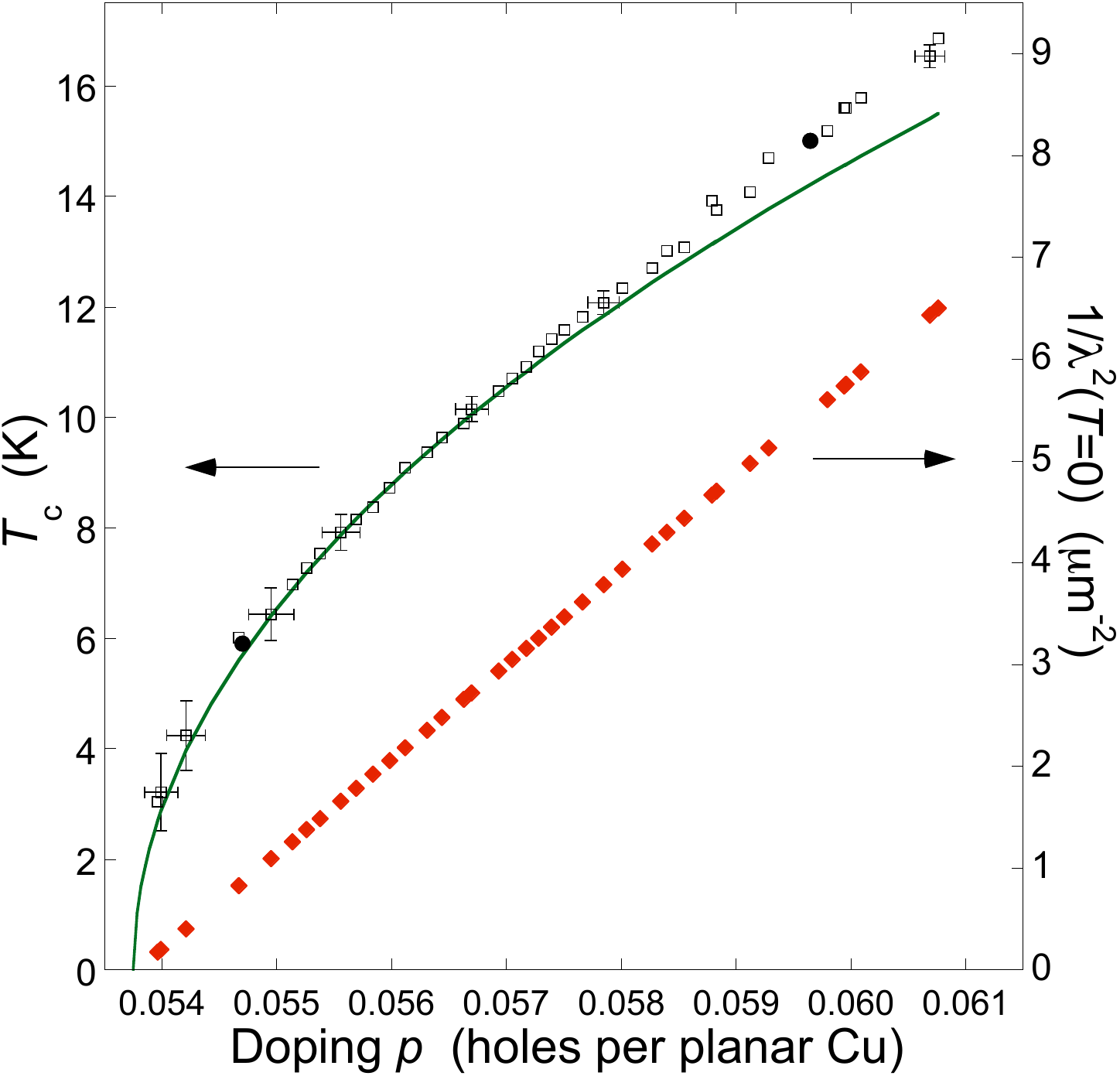}
\caption{\label{fig4} (color online).  The results of assuming a linear mapping between doping and $\rho_{\rm s0}$ (solid diamonds).  $T_{\rm c}$ (open squares) varies as $(p - p_{\rm c})^{1/2}$ as $p \rightarrow p_{\rm c}$, with the solid line a square-root $T_{\rm c}(p)$ curve for comparison.   $T_{\rm c}(p)$ data from Ref.~\onlinecite{liang06} (solid circles) are sparse in this range.   
 Vertical bars on the $T_{\rm c}$ data show transition widths estimated from rounding of $\sigma_1(T)$ fluctuation peaks.  The corresponding doping spreads (horizontal bars) show little variation with doping.}
\end{figure}

Previous studies of superfluid density in the underdoped cuprates have focused on the strong correlation between $T_{\rm c}$ and $\rho_{s0} \equiv 1/\lambda^2(T\rightarrow 0)$ \cite{uemura89,zuev05}.  Our new data, taken on a single, high purity crystal, explore the highly underdoped region in considerably more detail.  The strong correlation between $T_{\rm c}$ and $\rho_{s0}$ remains, but $\rho_{s0}$ is up to an order of magnitude larger than in the earlier work on thin films \cite{zuev05}.  As in the thin film study, $\rho_{s0}$ falls continuously to zero on underdoping and varies approximately quadratically with $T_{\rm c}$ at low doping.  Recent experiments on the CuO$_2$ plane doping state in \ybco\  have established a  mapping between $T_{\rm c}(y)$ and $p$, the hole concentration per planar Cu \cite{liang06}.  In Fig.~\ref{fig3}, a linear fit to the $T_{\rm c}(p)$ data from Ref.~\onlinecite{liang06} has been used to determine $p$ and plot $\rho_{\rm s0}(p)$.   At higher doping $\rho_{\rm s0}$ has a linear doping dependence. This behaviour appears to be very robust: an extrapolation of the linear fit in Fig.~\ref{fig3} passes within 5\% of the  $ab$-averaged superfluid density of Ortho-II \ybco\ \cite{peregbarnea04}.  To the extent that the linear extrapolation of $T_{\rm c}(p)$ holds,  $\rho_{s0}(p)$ varies approximately quadratically close to the onset of superconductivity.   However, as we will discuss below, this quadratic behaviour is difficult to understand theoretically.  Later on we will present an alternative proposal in which it is $\rho_{\rm s0}$, not $T_{\rm c}$, whose linear doping dependence  extends to the edge of the superconducting phase. 

\begin{figure*}[t] 
\centerline{\includegraphics[width=173mm]{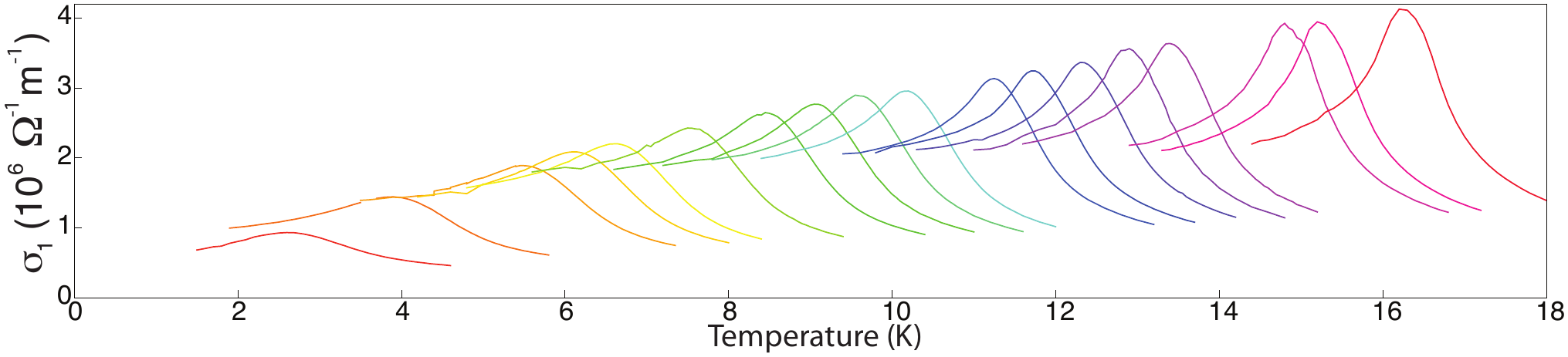}}
\caption{\label{fig5}  (color online). Fluctuation peaks in $\sigma_1(T)$ at 2.64~GHz as doping is varied.  Peak width is approximately constant in the higher doping range and then broadens considerably below $T_{\rm c} \approx 8$~K, consistent with a steepening of $T_{\rm c}(p)$ in that range.  }
\end{figure*}

The very low superfluid density, its continuous variation with doping, and the importance of 3D-$XY$ fluctuations near optimal doping \cite{kamal94, junod00, meingast01} together suggest that close to the critical doping, $p_{\rm c}$, the transition out of the superconducting state may be controlled by fluctuations near a quantum critical point.  In the scaling theory of a quantum phase transition, physical properties are related to a single, divergent correlation length, $\xi \propto |x|^{-\nu}$, where $x = (p - p_{\rm c})$.  Scaling analysis of the $XY$ model predicts $T_{\rm c} \propto \xi^{-z} \propto |x|^{\nu z}$ and $\rho_{\rm s0} \propto \xi^{-(d - 2 + z)} \propto |x|^{\nu(d - 2 + z)}$, where $z$ is the dynamical critical exponent \cite{sachdev99,herbut04,kopp05}. Together, these relations imply $T_{\rm c} \propto \rho_{s0}^{z/(d - 2 + z)}$, independent of $\nu$.  In $d=2$ this requires $T_{\rm c} \propto \rho_{\rm s0}$, whereas our experiments show $T_{\rm c} \sim \rho_{\rm s0}^{1/2}$.  In $d = 3$, $T_{\rm c} \propto \rho_{s0}^{z/(1 + z)}$ and is consistent with our observations if $z = 1$, the so-called $(3 + 1)$D-$XY$ universality class \cite{franz06}.  $D = 4$ is the upper critical dimension of the $XY$ model, so mean-field critical behaviour ($\nu = \frac{1}{2}$) should follow. However, this choice of $\nu$ is not compatible with the doping dependences of $T_{\rm c}$ and $\rho_{s0}$ shown in Fig.~\ref{fig3}: scaling arguments predict $T_{\rm c} \propto x^{1/2}$ and $\rho_{s0} \propto x$ whereas the analysis shown in Fig.~\ref{fig3} has $T_{\rm c} \propto x$ and therefore $\rho_{s0} \propto x^2$.  This difficulty may well stem from our determination of doping in Fig.~\ref{fig3}, which is based on a linear fit and extrapolation of the $T_{\rm c}(p)$ data from Ref.~\onlinecite{liang06}.    The sparseness of the $T_{\rm c}(p)$ data, along with the lack of an anchor point as $T_c \rightarrow 0$, allow a different interpretation in the low doping regime.  In Fig.~\ref{fig3}, all data with $T_{\rm c} \ge 8$~K follow an accurately linear doping dependence of $\rho_{\rm s0}$, which seems to extrapolate well to much higher dopings.  The robustness of this form leads us to put forward the following suggestion as a means of understanding the data.  We propose that $\rho_{s0}(p)$ is, in actual fact, proportional to $(p - p_{\rm c})$ in the low doping range and use this linear mapping to determine $T_{\rm c}(p)$.  The results of this analysis are shown in Fig.~\ref{fig4}.  The mapping leaves $T_{\rm c}(p)$ consistent with the data of Ref.~\onlinecite{liang06}; the effect is simply to refine the  $T_{\rm c}(p)$ curve in the vicinity of the critical doping. $T_{\rm c}(p)$ now grows initially as $(p - p_{\rm c})^{1/2}$, in accord with scaling arguments, before crossing over to a linear doping dependence at higher dopings. Over a substantial doping range this yields the well-known result that $T_{\rm c}$ scales approximately with $\rho_{\rm s0}$ \cite{uemura89}.  In addition, our proposed interpretation gives a plausible explanation of another aspect of the data.  At the lowest dopings, the  fluctuation peaks in $\sigma_1(T)$ broaden considerably, as shown in Figs.~\ref{fig4} and \ref{fig5}.  In the presence of small, macroscopic variations in oxygen concentration, such broadening would be a natural consequence of a steepening of $T_{\rm c}(p)$ on the approach to $p_{\rm c}$.  Future experiments, including those of the sort carried out in Ref.~\onlinecite{liang06}, will be needed to confirm our conjecture.  However, the proposed scenario has several compelling features, not least of which is a substantial simplification of our theoretical picture of the transition from nonsuperconductor to superconductor in the underdoped cuprates.

In summary, at the critical doping for superconductivity $\rho_{\rm s0}$ becomes nonzero and appears to grow linearly with doping, at a rate that remains constant up to much higher doping.  Fluctuations in the low doped range are three dimensional, with no indication of the physics of 2D vortex unbinding, and with critical exponents characteristic of $(3 + 1)$D-$XY$ universality.  Quantum fluctuations appear to control $T_{\rm c}$ in the critical region, with $T_{\rm c} \propto (p - p_{\rm c})^{1/2}$, but become less effective at depleting $\rho_{\rm s}$ away from $p_{\rm c}$, allowing $T_{\rm c}(p)$ to cross over to a linear doping dependence.

We acknowledge useful discussions with A.~J.~Berlinsky, M.~J.~Case, S.~Chakravarty, J.~Cooper, J.~C.~Davis, J.~S.~Dodge,  M.~Franz, I.~F.~Herbut, S.~Kivelson, J.~E.~Sonier and Z.~Te\u{s}anovi\'c. This work was funded by the National
Science and Engineering Research Council of Canada and the Canadian Institute for Advanced
Research.

%\bibliography{uYBCOSuperfluid2007}% Produces the bibliography via BibTeX.

\begin{thebibliography}{33}
\expandafter\ifx\csname natexlab\endcsname\relax\def\natexlab#1{#1}\fi
\expandafter\ifx\csname bibnamefont\endcsname\relax
  \def\bibnamefont#1{#1}\fi
\expandafter\ifx\csname bibfnamefont\endcsname\relax
  \def\bibfnamefont#1{#1}\fi
\expandafter\ifx\csname citenamefont\endcsname\relax
  \def\citenamefont#1{#1}\fi
\expandafter\ifx\csname url\endcsname\relax
  \def\url#1{\texttt{#1}}\fi
\expandafter\ifx\csname urlprefix\endcsname\relax\def\urlprefix{URL }\fi
\providecommand{\bibinfo}[2]{#2}
\providecommand{\eprint}[2][]{\url{#2}}

\bibitem[{\citenamefont{Orenstein and Millis}(2000)}]{orenstein00}
\bibinfo{author}{\bibfnamefont{J.}~\bibnamefont{Orenstein}} \bibnamefont{and}
  \bibinfo{author}{\bibfnamefont{A.~J.} \bibnamefont{Millis}},
  \bibinfo{journal}{Science} \textbf{\bibinfo{volume}{288}},
  \bibinfo{pages}{468} (\bibinfo{year}{2000}).

\bibitem[{\citenamefont{Emery and Kivelson}(1995)}]{emery95}
\bibinfo{author}{\bibfnamefont{V.~J.} \bibnamefont{Emery}} \bibnamefont{and}
  \bibinfo{author}{\bibfnamefont{S.~A.} \bibnamefont{Kivelson}},
  \bibinfo{journal}{Nature} \textbf{\bibinfo{volume}{374}},
  \bibinfo{pages}{434} (\bibinfo{year}{1995}).

\bibitem[{\citenamefont{Franz and Te\u{s}anovi\'c}(2001)}]{franz01}
\bibinfo{author}{\bibfnamefont{M.}~\bibnamefont{Franz}} \bibnamefont{and}
  \bibinfo{author}{\bibfnamefont{Z.}~\bibnamefont{Te\u{s}anovi\'c}},
  \bibinfo{journal}{Phys.\ Rev.\ Lett.} \textbf{\bibinfo{volume}{87}},
  \bibinfo{pages}{257003} (\bibinfo{year}{2001}); \bibinfo{author}{\bibfnamefont{M.}~\bibnamefont{Franz}},
  \bibinfo{author}{\bibfnamefont{Z.}~\bibnamefont{Te\u{s}anovi\'c}},
  \bibnamefont{and} \bibinfo{author}{\bibfnamefont{O.}~\bibnamefont{Vafek}},
  \bibinfo{journal}{Phys.\ Rev.\ B} \textbf{\bibinfo{volume}{66}},
  \bibinfo{pages}{054535} (\bibinfo{year}{2002}).

%\bibitem[{\citenamefont{Franz et~al.}(2002)\citenamefont{Franz,
%  Te\u{s}anovi\'c, and Vafek}}]{franz02}
%\bibinfo{author}{\bibfnamefont{M.}~\bibnamefont{Franz}},
%  \bibinfo{author}{\bibfnamefont{Z.}~\bibnamefont{Te\u{s}anovi\'c}},
%  \bibnamefont{and} \bibinfo{author}{\bibfnamefont{O.}~\bibnamefont{Vafek}},
%  \bibinfo{journal}{Phys.\ Rev.\ B} \textbf{\bibinfo{volume}{66}},
%  \bibinfo{pages}{054535} (\bibinfo{year}{2002}).

\bibitem[{\citenamefont{Herbut}(2002{\natexlab{a}})}]{herbut02}
\bibinfo{author}{\bibfnamefont{I.~F.} \bibnamefont{Herbut}},
  \bibinfo{journal}{Phys.\ Rev.\ Lett.} \textbf{\bibinfo{volume}{88}},
  \bibinfo{pages}{047006} (\bibinfo{year}{2002}{\natexlab{a}}).

\bibitem[{\citenamefont{Herbut}(2002{\natexlab{b}})}]{herbut02a}
\bibinfo{author}{\bibfnamefont{I.~F.} \bibnamefont{Herbut}},
  \bibinfo{journal}{Phys.\ Rev.\ B} \textbf{\bibinfo{volume}{66}},
  \bibinfo{pages}{094504} (\bibinfo{year}{2002}{\natexlab{b}}).

\bibitem[{\citenamefont{Herbut and Case}(2004)}]{herbut04}
\bibinfo{author}{\bibfnamefont{I.~F.} \bibnamefont{Herbut}} \bibnamefont{and}
  \bibinfo{author}{\bibfnamefont{M.~J.} \bibnamefont{Case}},
  \bibinfo{journal}{Phys.\ Rev.\ B} \textbf{\bibinfo{volume}{70}},
  \bibinfo{pages}{094516} (\bibinfo{year}{2004}).

\bibitem[{\citenamefont{Herbut}(2005)}]{herbut05}
\bibinfo{author}{\bibfnamefont{I.~F.} \bibnamefont{Herbut}},
  \bibinfo{journal}{Phys.\ Rev.\ Lett.} \textbf{\bibinfo{volume}{94}},
  \bibinfo{pages}{237001} (\bibinfo{year}{2005}).

\bibitem[{\citenamefont{Franz and Iyengar}(2006)}]{franz06}
\bibinfo{author}{\bibfnamefont{M.}~\bibnamefont{Franz}} \bibnamefont{and}
  \bibinfo{author}{\bibfnamefont{A.~P.} \bibnamefont{Iyengar}},
  \bibinfo{journal}{Phys.\ Rev.\ Lett.} \textbf{\bibinfo{volume}{96}},
  \bibinfo{pages}{047007} (\bibinfo{year}{2006}).

\bibitem[{\citenamefont{Uemura et~al.}(1989)}]{uemura89}
\bibinfo{author}{\bibfnamefont{Y.~J.} \bibnamefont{Uemura}}
  \bibnamefont{et~al.}, \bibinfo{journal}{Phys.\ Rev.\ Lett.}
  \textbf{\bibinfo{volume}{62}}, \bibinfo{pages}{2317} (\bibinfo{year}{1989}).

\bibitem[{\citenamefont{Corson et~al.}(1999)}]{corson99}
\bibinfo{author}{\bibfnamefont{J.}~\bibnamefont{Corson}} \bibnamefont{et~al.},
  \bibinfo{journal}{Nature} \textbf{\bibinfo{volume}{398}},
  \bibinfo{pages}{221} (\bibinfo{year}{1999}).

\bibitem[{\citenamefont{Wang et~al.}(2003)}]{wang03}
\bibinfo{author}{\bibfnamefont{Y.}~\bibnamefont{Wang}} \bibnamefont{et~al.},
  \bibinfo{journal}{Science} \textbf{\bibinfo{volume}{299}},
  \bibinfo{pages}{86} (\bibinfo{year}{2003}).

\bibitem[{\citenamefont{Liang et~al.}(2002)}]{liang02}
\bibinfo{author}{\bibfnamefont{R.}~\bibnamefont{Liang}} \bibnamefont{et~al.},
  \bibinfo{journal}{Physica C} \textbf{\bibinfo{volume}{383}},
  \bibinfo{pages}{1} (\bibinfo{year}{2002}).

\bibitem[{\citenamefont{Liang et~al.}(2006)\citenamefont{Liang, Bonn, and
  Hardy}}]{liang06}
\bibinfo{author}{\bibfnamefont{R.}~\bibnamefont{Liang}},
  \bibinfo{author}{\bibfnamefont{D.~A.} \bibnamefont{Bonn}}, \bibnamefont{and}
  \bibinfo{author}{\bibfnamefont{W.~N.} \bibnamefont{Hardy}},
  \bibinfo{journal}{Phys.\ Rev.\ B} \textbf{\bibinfo{volume}{73}},
  \bibinfo{pages}{180505} (\bibinfo{year}{2006}).

\bibitem[{\citenamefont{Hosseini et~al.}(2004)}]{hosseini04}
\bibinfo{author}{\bibfnamefont{A.}~\bibnamefont{Hosseini}}
  \bibnamefont{et~al.}, \bibinfo{journal}{Phys.\ Rev.\ Lett.}
  \textbf{\bibinfo{volume}{93}}, \bibinfo{pages}{107003}
  (\bibinfo{year}{2004}).

\bibitem[{\citenamefont{Zaanen et~al.}(1988)}]{zaanen88}
\bibinfo{author}{\bibfnamefont{J.}~\bibnamefont{Zaanen}} \bibnamefont{et~al.},
  \bibinfo{journal}{Phys.\ Rev.\ Lett.} \textbf{\bibinfo{volume}{60}},
  \bibinfo{pages}{2685} (\bibinfo{year}{1988}).

\bibitem[{\citenamefont{Veal et~al.}(1990)}]{veal90}
\bibinfo{author}{\bibfnamefont{B.~W.} \bibnamefont{Veal}} \bibnamefont{et~al.},
  \bibinfo{journal}{Phys.\ Rev.\ B} \textbf{\bibinfo{volume}{42}},
  \bibinfo{pages}{6305} (\bibinfo{year}{1990}).

\bibitem[{\citenamefont{Liang et~al.}(1998)\citenamefont{Liang, Bonn, and
  Hardy}}]{liang98}
\bibinfo{author}{\bibfnamefont{R.}~\bibnamefont{Liang}},
  \bibinfo{author}{\bibfnamefont{D.~A.} \bibnamefont{Bonn}}, \bibnamefont{and}
  \bibinfo{author}{\bibfnamefont{W.~N.} \bibnamefont{Hardy}},
  \bibinfo{journal}{Physica C} \textbf{\bibinfo{volume}{304}},
  \bibinfo{pages}{105} (\bibinfo{year}{1998}).

\bibitem[{\citenamefont{Huttema et~al.}(2006)}]{huttema05}
\bibinfo{author}{\bibfnamefont{W.~A.} \bibnamefont{Huttema}}
  \bibnamefont{et~al.}, \bibinfo{journal}{Rev.\ Sci.\ Instrum.}
  \textbf{\bibinfo{volume}{77}}, \bibinfo{pages}{023901}
  (\bibinfo{year}{2006}).

\bibitem[{\citenamefont{Fisher et~al.}(1991)\citenamefont{Fisher, Fisher, and
  Huse}}]{fisher91}
\bibinfo{author}{\bibfnamefont{D.~S.} \bibnamefont{Fisher}},
  \bibinfo{author}{\bibfnamefont{M.~P.~A.} \bibnamefont{Fisher}},
  \bibnamefont{and} \bibinfo{author}{\bibfnamefont{D.~A.} \bibnamefont{Huse}},
  \bibinfo{journal}{Phys.\ Rev.\ B} \textbf{\bibinfo{volume}{43}},
  \bibinfo{pages}{130} (\bibinfo{year}{1991}).

\bibitem[{\citenamefont{Bonn et~al.}(1994)}]{bonn94}
\bibinfo{author}{\bibfnamefont{D.~A.} \bibnamefont{Bonn}} \bibnamefont{et~al.},
  \bibinfo{journal}{Phys.\ Rev.\ B} \textbf{\bibinfo{volume}{50}},
  \bibinfo{pages}{4051} (\bibinfo{year}{1994}).

\bibitem[{\citenamefont{Hirschfeld and Goldenfeld}(1993)}]{hirschfeld93}
\bibinfo{author}{\bibfnamefont{P.~J.} \bibnamefont{Hirschfeld}}
  \bibnamefont{and}
  \bibinfo{author}{\bibfnamefont{N.}~\bibnamefont{Goldenfeld}},
  \bibinfo{journal}{Phys.\ Rev.\ B} \textbf{\bibinfo{volume}{48}},
  \bibinfo{pages}{R4219} (\bibinfo{year}{1993}); \bibinfo{author}{\bibfnamefont{P.~J.} \bibnamefont{Hirschfeld}},
  \bibinfo{author}{\bibfnamefont{W.~O.} \bibnamefont{Putikka}},
  \bibnamefont{and} \bibinfo{author}{\bibfnamefont{D.~J.}
  \bibnamefont{Scalapino}}, \bibinfo{journal}{Phys.\ Rev.\ B}
  \textbf{\bibinfo{volume}{50}}, \bibinfo{pages}{10250} (\bibinfo{year}{1994}).

%\bibitem[{\citenamefont{Hirschfeld et~al.}(1994)\citenamefont{Hirschfeld,
%  Putikka, and Scalapino}}]{hirschfeld94}
%\bibinfo{author}{\bibfnamefont{P.~J.} \bibnamefont{Hirschfeld}},
%  \bibinfo{author}{\bibfnamefont{W.~O.} \bibnamefont{Putikka}},
%  \bibnamefont{and} \bibinfo{author}{\bibfnamefont{D.~J.}
%  \bibnamefont{Scalapino}}, \bibinfo{journal}{Phys.\ Rev.\ B}
%  \textbf{\bibinfo{volume}{50}}, \bibinfo{pages}{10250} (\bibinfo{year}{1994}).

\bibitem[{\citenamefont{Pereg-Barnea et~al.}(2004)}]{peregbarnea04}
\bibinfo{author}{\bibfnamefont{T.}~\bibnamefont{Pereg-Barnea}}
  \bibnamefont{et~al.}, \bibinfo{journal}{Phys.\ Rev.\ B}
  \textbf{\bibinfo{volume}{69}}, \bibinfo{pages}{184513}
  (\bibinfo{year}{2004}).

\bibitem[{\citenamefont{Kamal et~al.}(1994)}]{kamal94}
\bibinfo{author}{\bibfnamefont{S.}~\bibnamefont{Kamal}} \bibnamefont{et~al.},
  \bibinfo{journal}{Phys. Rev. Lett.} \textbf{\bibinfo{volume}{73}},
  \bibinfo{pages}{1845} (\bibinfo{year}{1994}).

\bibitem[{\citenamefont{Junod et~al.}(2000)}]{junod00}
\bibinfo{author}{\bibfnamefont{A.}~\bibnamefont{Junod}} \bibnamefont{et~al.},
  \bibinfo{journal}{Physica B} \textbf{\bibinfo{volume}{280}},
  \bibinfo{pages}{214} (\bibinfo{year}{2000}).

\bibitem[{\citenamefont{Meingast et~al.}(2001)}]{meingast01}
\bibinfo{author}{\bibfnamefont{C.}~\bibnamefont{Meingast}}
  \bibnamefont{et~al.}, \bibinfo{journal}{Phys. Rev. Lett}
  \textbf{\bibinfo{volume}{86}}, \bibinfo{pages}{1606} (\bibinfo{year}{2001}).

\bibitem[{\citenamefont{Ioffe and Millis}(2002)}]{ioffe02}
\bibinfo{author}{\bibfnamefont{L.~B.} \bibnamefont{Ioffe}} \bibnamefont{and}
  \bibinfo{author}{\bibfnamefont{A.~J.} \bibnamefont{Millis}},
  \bibinfo{journal}{J.\ Phys.\ Chem.\ Solids} \textbf{\bibinfo{volume}{63}},
  \bibinfo{pages}{2259} (\bibinfo{year}{2002}).

\bibitem[{\citenamefont{Zuev et~al.}()}]{zuev04}
\bibinfo{author}{\bibfnamefont{Y.}~\bibnamefont{Zuev}} \bibnamefont{et~al.},
  \emph{\bibinfo{title}{unpublished}}, \eprint{cond-mat/0407113}; \bibinfo{author}{\bibfnamefont{I.}~\bibnamefont{Hetel}},
  \bibinfo{author}{\bibfnamefont{T.~R.} \bibnamefont{Lemberger}},
  \bibnamefont{and} \bibinfo{author}{\bibfnamefont{M.}~\bibnamefont{Randeria}},
  \bibinfo{journal}{Nature Phys.} \textbf{\bibinfo{volume}{3}},
  \bibinfo{pages}{700} (\bibinfo{year}{2007}).

%\bibitem[{\citenamefont{Hetel et~al.}(2007)\citenamefont{Hetel, Lemberger, and
%  Randeria}}]{hetel07}
%\bibinfo{author}{\bibfnamefont{I.}~\bibnamefont{Hetel}},
%  \bibinfo{author}{\bibfnamefont{T.~R.} \bibnamefont{Lemberger}},
%  \bibnamefont{and} \bibinfo{author}{\bibfnamefont{M.}~\bibnamefont{Randeria}},
%  \bibinfo{journal}{Nature Phys.} \textbf{\bibinfo{volume}{3}},
%  \bibinfo{pages}{700} (\bibinfo{year}{2007}).

\bibitem[{\citenamefont{Zuev et~al.}(2005)\citenamefont{Zuev, Kim, and
  Lemberger}}]{zuev05}
\bibinfo{author}{\bibfnamefont{Y.}~\bibnamefont{Zuev}},
  \bibinfo{author}{\bibfnamefont{M.~S.} \bibnamefont{Kim}}, \bibnamefont{and}
  \bibinfo{author}{\bibfnamefont{T.~R.} \bibnamefont{Lemberger}},
  \bibinfo{journal}{Phys.\ Rev.\ Lett.} \textbf{\bibinfo{volume}{95}},
  \bibinfo{pages}{137002} (\bibinfo{year}{2005}).

\bibitem[{\citenamefont{Kopp and Chakravarty}(2005)}]{kopp05}
\bibinfo{author}{\bibfnamefont{A.}~\bibnamefont{Kopp}} \bibnamefont{and}
  \bibinfo{author}{\bibfnamefont{S.}~\bibnamefont{Chakravarty}},
  \bibinfo{journal}{Nature Phys.} \textbf{\bibinfo{volume}{1}},
  \bibinfo{pages}{53} (\bibinfo{year}{2005}).

\bibitem[{\citenamefont{Sachdev}(1999)}]{sachdev99}
\bibinfo{author}{\bibfnamefont{S.}~\bibnamefont{Sachdev}},
  \emph{\bibinfo{title}{Quantum Phase Transitions}}
  (\bibinfo{publisher}{Cambridge University Press}, \bibinfo{year}{1999}).

\end{thebibliography}

\end{document}